\documentclass[10pt,journal,compsoc]{IEEEtran}
\usepackage{array}

\ifCLASSOPTIONcompsoc
  \usepackage[nocompress]{cite}
\else
  \usepackage{cite}
\fi

\ifCLASSINFOpdf
  \usepackage[pdftex]{graphicx}
  \graphicspath{{../pdf/}{../jpeg/}}
  \DeclareGraphicsExtensions{.pdf,.jpeg,.png}
\else
  \usepackage{graphicx}                
  \DeclareGraphicsExtensions{.eps}     
\fi
\usepackage{fixltx2e}

\usepackage{stfloats}
\ifCLASSOPTIONcaptionsoff
 \usepackage[nomarkers]{endfloat}
\let\MYoriglatexcaption\caption
\renewcommand{\caption}[2][\relax]{\MYoriglatexcaption[#2]{#2}}
\fi
\usepackage{url}

\usepackage{color}
\usepackage{amsmath}

\ifCLASSOPTIONcompsoc
 \usepackage[caption=false,font=footnotesize,labelfont=sf,textfont=sf,captionskip=1pt]{subfig}
\else
 \usepackage[caption=false,font=footnotesize]{subfig}
\fi

\usepackage[hidelinks]{hyperref}
\usepackage{amssymb}
\usepackage{amsfonts}
\usepackage{floatrow}
\usepackage{bm}
\usepackage[locale=FR]{siunitx}
\usepackage[font=footnotesize,labelfont=sf,textfont=sf]{caption}
\usepackage{tikz}
\usepackage{pgfplots}

\usepackage{multicol}
\usepackage{paralist}
\usepackage[mathscr]{euscript}

\newcommand{\norm}[1]{\left\lVert#1\right\rVert}
\newcommand{\mathrmbf}[1]{\mathrm{\mathbf{#1}}}
\DeclareMathAlphabet{\mathcal}{OMS}{cmsy}{m}{n}
\SetMathAlphabet{\mathcal}{bold}{OMS}{cmsy}{b}{n}
\newcommand{\bigO}{\mathcal{O}}

\usepackage[keeplastbox]{flushend}

\begin{document}
%
\title{A Visual Analytics Approach for Hardware System Monitoring with Streaming Functional Data Analysis}

%
%
%
%

\author{Shilpika,
        Takanori~Fujiwara,
        Naohisa~Sakamoto,
        Jorji~Nonaka,
        and~Kwan-Liu~Ma
\IEEEcompsocitemizethanks{\IEEEcompsocthanksitem Shilpika, T. Fujiwara, K.-L. Ma are with University of California, Davis.\protect\\
    E-mail: \{fshilpika, tfujiwara, klma\}@ucdavis.edu.
\IEEEcompsocthanksitem N. Sakamoto is with Kobe University. \protect\\
E-mail: naohisa.sakamoto@people.kobe-u.ac.jp.
\IEEEcompsocthanksitem J. Nonaka is with RIKEN R-CCS. E-mail: jorji@riken.jp.
}
\thanks{Manuscript received October 22, 2021; revised February 18, 2022.}}

%
%

\markboth{ }
{Shell \MakeLowercase{\textit{et al.}}: A Visual Analytics Approach for Hardware System Monitoring with Streaming Functional Data Analysis}


\def\markboth#1#2{\def\leftmark{#1}\def\rightmark{#2}}
\markboth{\tiny{To appear in IEEE Transactions on Visualization and Computer Graphics. Previously, titled ``A Visual Analytics Approach to Monitor Time-Series Data with Incremental and Progressive Functional Data Analysis.''}}{}

%



\IEEEtitleabstractindextext{%
\begin{abstract}
Many real-world applications involve analyzing time-dependent phenomena, which are intrinsically functional, consisting of curves varying over a continuum (e.g., time).
When analyzing continuous data, functional data analysis (FDA) provides substantial benefits, such as the ability to study the derivatives and to restrict the ordering of data. 
However, continuous data inherently has infinite dimensions, and for a long time series, FDA methods often suffer from high computational costs. 
The analysis problem becomes even more challenging when updating the FDA results for continuously arriving data.
In this paper, we present a visual analytics approach for monitoring and reviewing time series data streamed from a hardware system with a focus on identifying outliers by using FDA.
To perform FDA while addressing the computational problem, we introduce new incremental and progressive algorithms that promptly generate the magnitude-shape (MS) plot, which conveys both the functional magnitude and shape outlyingness of time series data.
In addition, by using an MS plot in conjunction with an FDA version of principal component analysis, we enhance the analyst's ability to investigate the visually-identified outliers. 
We illustrate the effectiveness of our approach with two use scenarios using real-world datasets. The resulting tool is evaluated by industry experts using real-world streaming datasets.
\end{abstract}

}

\maketitle


\IEEEdisplaynontitleabstractindextext

%
\IEEEpeerreviewmaketitle

\IEEEraisesectionheading{\section{Introduction}\label{sec:introduction}}

\IEEEPARstart{T}{o} ensure normal operation and adequate performance
of hardware systems, such as those in an assembly plant
or a supercomputer center, various monitoring mechanisms examine all aspects of the system's data collected at high frequency in real-time~\cite{8514917}.
The ability to instantaneously process and analyze the resulting time series data thus becomes 
pertinent to examining various underlying system behaviors in order to detect and promptly react to system failures or inefficiencies. 

High-velocity time series data is intrinsically functional as it can be represented in the form of curves or surfaces with weak assumptions 
of smoothness being permitted~\cite{wangdavis}.
Functional data analysis (FDA) is a branch of statistics for analyzing such data~\cite{ramsaysilverman}. 
FDA incorporates  statistical methodologies to capture the underlying properties and structure of the data.
The statistical methods and models in FDA are typically presented as continuous functions.
A wide range of methods has been developed to handle functional data, such as functional principal component analysis (FPCA), functional regression models, and functional canonical correlation analysis~\cite{ramsaysilverman}.
As these names indicate, many FDA methods resemble those developed for conventional discrete analysis. 
Because FDA handles data with continuous functions, FDA has advantages in studying the derivatives of the data and maintaining temporal data ordering~\cite{wangdavis}. 
FDA is beneficial when dealing with high-frequency mission-critical streaming data, where we want to analyze the changes in the data behavior with time while also viewing the past behavior. Recently, FDA has seen growth with time-critical applications in various fields, such as biology, medicine, finance, and engineering~\cite{Millan-Roures2018,8819420}.

While FDA methods provide capabilities for analyzing data over a continuum like time series data, they often suffer from high computational costs, which increase with the number of time points.
This can be a critical problem with real-world monitoring applications as they keep generating new time points, resulting in infinitely long time series. 

With this work, we aim to reduce the computational overheads of using  FDA  on streaming time series data 
while retaining the capabilities of the in-depth analysis provided by FDA. 
To do so, we introduce a visual analytics approach for 
continuously processing time series data that grows over time with a focus on identifying outliers by using FDA. 
In particular, we design new incremental and progressive  algorithms that promptly generate the magnitude-shape (MS) plot~\cite{mvod}, 
which reveals the outlyingness of both the functional magnitude and shape (or variations) of the time series. 
These updates utilizing the previously computed results 
address the computational overhead introduced by handling growing time series data in bulk. 
Also, with the support of FPCA, our approach helps analysts investigate the visually-identified outliers from the MS plot.

The main contributions of our work are to provide:
\begin{compactitem}
    \item new incremental and progressive algorithms to generate the MS plot, which enables analyses of time series data collected from online data streams; 
    \item augmentation of analysis using the MS plot with FPCA and interactive visualization to aid in reviewing clusters identified from the MS plot;
    \item two use scenarios on real-world datasets demonstrating the effectiveness of our approach, both of which involve the addition of (1) new time points and (2) new independent time series.
\end{compactitem}

\section{Background and Related Work}

We discuss relevant works in streaming data visualization and provide a background to FDA, including FPCA and the MS plot~\cite{mvod}---FDA methods utilized in our work.

\subsection{Streaming Data Visualization}

Visualization of streaming data is a challenging task since visualizations need to be continually updated with incoming data. 
The major bottlenecks in visualizing continuous data streams are computational cost and cognitive load. 
As for the cognitive load, Krstajic and Keim~\cite{krstajic2013} discussed the trade-offs between updating a view when a new data point is fed, which may lead to loss of mental map, or not updating a view, which may lead to loss of information.
A comprehensive survey by Dasgupta et al.~\cite{dasgupta2017} further characterized challenges in perception and cognition of streaming visualizations. While our work considers the cognitive load of visualizations, we mainly address the computational cost when producing visualizations.

Past works on incremental algorithms have visualized up-to-date results while avoiding rising computation costs.
For example, Tanahashi et al.~\cite{tanahashi2015efficient} built a streaming storyline visualization. 
Liu et al.~\cite{liu2016online} introduced a streaming tree cut algorithm to visualize incoming topics from text streams.
Crnovrsanin et al.~\cite{crnovrsanin2017incremental} developed a GPU accelerated incremental force-directed layout. 
Other researchers developed incremental dimensionality reduction (DR)~\cite{tak2019} and change-point detection methods~\cite{kesavan2020visual}.
Katragadda et al.~\cite{vastream} developed a visual analytics tool for high-velocity streaming data using hardware and software sandbox environments.

Another potential approach for streaming data analysis is progressive visual analytics~\cite{turkay2018progressive}, which provides reasonable intermediate results within a required latency when the computational cost for an entire calculation is too high. 
This latency requirement is common with streaming data visualizations. Thus, for streaming visualization, we can utilize progressive algorithms, such as the progressive version of t-SNE~\cite{pezzotti2017approximated}, UMAP~\cite{pumap}, and time series clustering~\cite{kesavan2020visual}.

Compared to the works above, our work addresses the problem of using FDA methods to generate visualizations from data streams.
Our algorithm demonstrates a new capability for visual analytics of functional data (including time series data) in a streaming, timely manner.

\subsection{Functional Data Analysis (FDA)}

Functional data is data procured from continuous phenomena of space or time and represented in the form of smooth functions.
For example, sensor readings at components in a supercomputer can be viewed as data produced from a function $\mathrmbf{X}$ of $K$ variables (e.g., variables corresponding to temperature and voltage).
Then, $\mathrmbf{X}_n(t)$ is an observation (e.g., temperature) corresponding to a machine component $n$ (e.g., one CPU temperature sensor) at time $t$.
While readings are collected at finite resolutions in practice, the temperature, for example, continuously exists over time. Thus, it is natural to model and analyze $\mathrmbf{X}_n(t)$ with a continuous function defined over time from the observations.

The field of FDA has seen advances in functional data modeling, clustering, differential analysis, and outlier detection. Ferraty and Vieu~\cite{ferView} popularized the functional non-parametric statistics with free-modeling. Ramsay and Silverman~\cite{ramsaysilverman} applied parametric statistics, such as linear regression and principal components analysis, to the functional domain. 
Horvath and Kokoszka~\cite{horKok} developed statistical methods for inference on functional data. 
FDA intrinsically maintains the data order and captures the data evolution of a function over a continuum. 
FDA allows for analysis on the rates of changes, derivatives of functions, among others; thus, FDA can provide various insights that are difficult to find with other analyses.
Since functional data consists of smooth, continuous sequences collected from various sources, outlier detection is often performed as a preliminary step in analysis to distinguish patterns between sequences. 
In FDA, the concept of statistical \textit{depth} plays a principal role in identifying outliers. 
Generally, statistical depth is used to measure how far a sample is from a median of multivariate data as the median is located at the deepest point as per this concept~\cite{mosler2013depth}.  
Since the introduction by Tukey~\cite{Tukey}, various popular depth measures have been developed, which include half-space depth~\cite{Tukey}, projection depth~\cite{zuo2004stahel}, and spatial depth~\cite{vardi}. 

Several visualization tools have been developed for FDA to effectively communicate underlying characteristics otherwise not apparent through summary statistics and models. 
Hyndman and Shang~\cite{HyndmanShang} introduced tools to visualize large amounts of functional data, such as functional versions of bagplots (multidimensional boxplots) and highest-density-region plots. 
Other popular tools include functional boxplots~\cite{Tukey} and MS plots~\cite{mvod}. 
The MS plot is designed for the visual identification of outliers using the magnitude and shape outlyingness of each series of functional data. 
With simulated and real-world examples, the authors have shown how the MS plot is superior in identifying potential outliers. 
We enhance the MS plot for streaming data analysis by providing incremental and progressive algorithms to enable timely updates. 
Our algorithm can produce identical results with ordinary MS plots when records are added along the continuum (e.g., time points addition). 
When adding new series (e.g., time series), our algorithm performs progressive updates while maintaining a  reasonable computation error. 

FPCA is one of the most popular FDA methods.
It captures the principal directions of modes of variations and performs DR. 
Using the basis spanned by the principal component, FPCA can summarize subject-specific features.
Karhunen~\cite{karhunen1946spektraltheorie} and Loeve~\cite{loeve}
made the first advancements in FPCA through optimal data expansion of a continuous stochastic process.
Building on this work, Rao~\cite{Rao1958} performed statistical tests comparing temporal growth curves. 
Extensions of FPCA are listed in the survey by Shang~\cite{shang2014}.  
Practical applications of FPCA include analyses of functional magnetic resonance imaging~\cite{Viviani}, age-specific mortality rating~\cite{ullah}, and income density curves~\cite{Kneip}. Since it is a broadly used and perfected FDA method, our tool uses the FPCA in conjunction with the MS plot.
Using the MS plot, the user can select functional data of interest and analyze it further with FPCA. Thus, we harness the benefits of identifying functional trends and patterns using FPCA for large data streams from the promptly generated MS plot.

\section{Methodology}
\label{sec:methodology}

\begin{figure}[tb]
	\centering
 	\captionsetup{farskip=0pt}
    \includegraphics[width=1.0\linewidth]{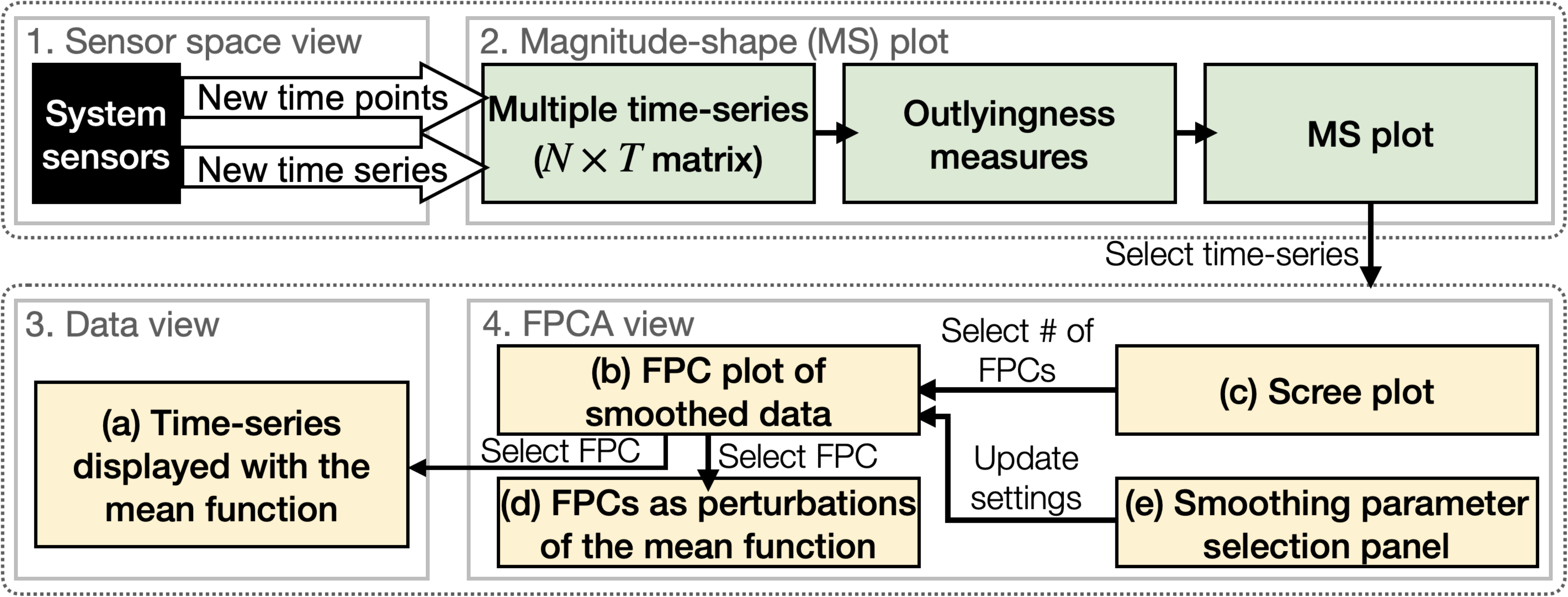}
    \caption{General organization of our visualization tool. 
    }
	\label{fig:architecture}
\end{figure}

This section describes the organization of our visual analytics tool and the back-end analysis methods.

\subsection{Overall Organization}\label{sec:overallO}

\autoref{fig:architecture} shows the organization of our visual analytics tool. 
The tool can be divided into two main components: (1) generation of the MS plot from streaming feed (top) and (2) interactive analysis over the generated MS plot with the auxiliary visualizations from other FDA methods, including FPCA and smoothing functions (bottom).

To make explanations concise and concrete, we describe our algorithms with streaming (multivariate) time series data as an example of functional data. 
While such data is our main analysis target, our approach can be applied to other types of functional data as we use general FDA methods such as the MS plot and FPCA. 
Here we denote the numbers of existing time points, variables (e.g., measured temperature and voltage), and time series in the current multivariate time series data with $T$, $K$, and $N$, respectively. 

Time series data is collected from various sensors housed at various levels of the hierarchy within a system. If the sensor system topology is available and can be visualized, we have an optional sensor space view (\autoref{fig:architecture}-1) where we display the monitoring system components and highlight the individual components from which the measurements/readings are being processed. 
In \autoref{fig:architecture}-2, we first initialize the magnitude and shape outlyingness measures from the current set of functional data.
Then, as shown in \autoref{fig:architecture}-1, data updates from the stream can be the addition of either time point (i.e., $T \! \to \! T \! + \! 1$) or time series ($N \! \to \! N \! + \! 1$).

Based on the updates, we compute the directional outlyingness measures~\cite{mvod} incrementally without any approximations for the addition of a time point and progressively with approximations for the addition of a time series (\autoref{fig:architecture}-2).
The progressive update introduces errors when compared to the actual results. 
If the errors exceed a predefined threshold, the results are recomputed for all the selected time series in the back-end and made available in the UI upon completion. 
We make sure that incremental and progressive updates do not co-occur to avoid overwriting results. 
The MS plot is updated accordingly.
An example of the MS plot can be seen in \autoref{fig:msplot}-b.
By default, we update the MS plot every 10 addition of time points (e.g., 5 seconds when a stream rate is every $0.5$ second) to help analysts maintain their mental map while avoiding large information loss.

In the MS plot, each \textit{circle} represents one time series. 
The user can select multiple time series to visualize them in the data view (\autoref{fig:architecture}-3).
The selected time series are also processed by 
the FPCA pipeline (\autoref{fig:architecture}-4), and the results are shown in 
the functional principal components (FPCs) plot, the scree plot~\cite{scree}, 
and the plot showing FPCs as the perturbation to the mean function. 
In addition, the UI includes a panel for the user to select smoothing basis functions 
for applying FPCA. 
\autoref{sec:fpca_and_visualization} describes the detail of each of these plots.

\subsection{Incremental and Progressive Generation of the Magnitude-Shape Plot}
\label{sec:inc_prog_msplot}

We describe our algorithms\footnote{The related source code is available at \url{https://github.com/sshilpika/streaming-ms-plot}.} and system implementations that are designed to generate the MS plot from data streams.

\subsubsection{Magnitude-Shape (MS) Plot}
\label{sec:msplot}

\begin{figure}[tb]
	\centering
 	\captionsetup{farskip=0pt}
    \includegraphics[width=1.0\linewidth]{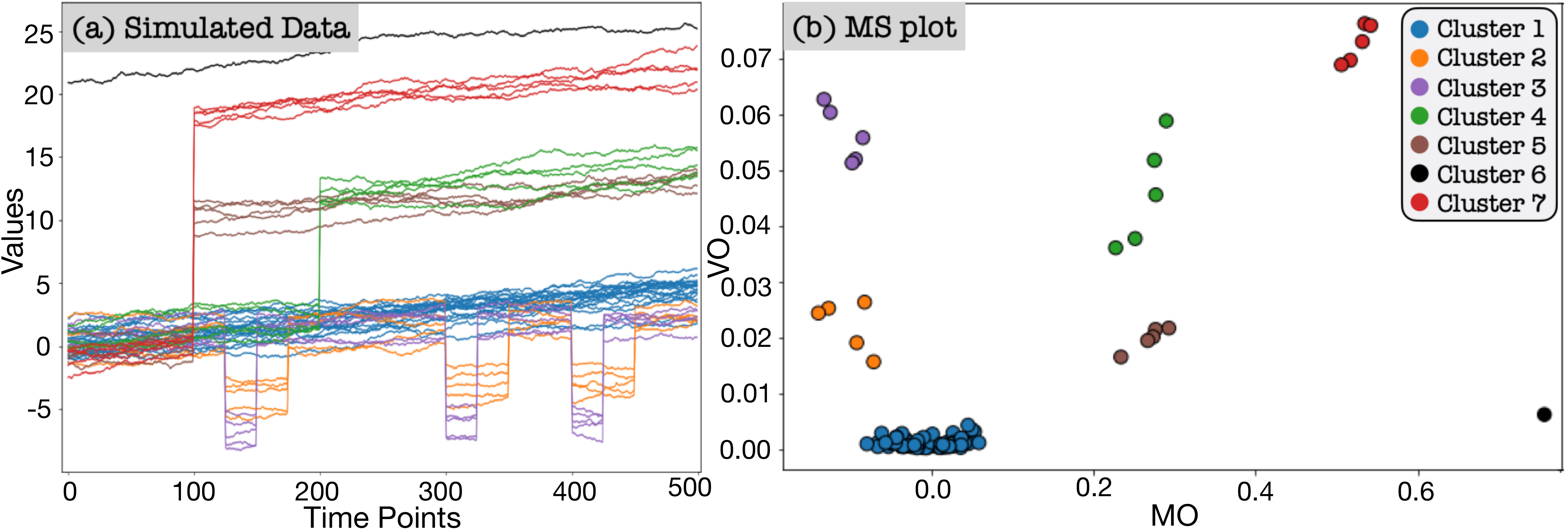}
    \caption{Visual outlier detection with the MS plot: (a) the simulated functional data and (b) the MS plot with magnitude outlyingness (MO) and variational outlyingness (VO) along $x$-, and $y$-axes, respectively.}
	\label{fig:msplot}
\end{figure}

We briefly introduce the MS plot~\cite{mvod}, which we use as a primary visual analysis tool, and extend it to use in a streaming setting.
The MS plot is a scatterplot that shows, for each time series, the corresponding outlyingness measures: mean directional outlyingness and variation of directional outlyingness~\cite{dodai}, for example, as $x$- and $y$-coordinates, respectively. 
As shown in \autoref{fig:msplot}, the MS plot depicts how much a time series has a different magnitude and shape with the other time series, and thus visually reveals outliers.  

A time series can be considered an outlier if it behaves in a manner inconsistent with the other time series. 
One of the measures of the outlyingness of time series is the Stahel-Donoho outlyingness~\cite{zuo2004stahel}. 
This measure is suited when values in time series data roughly follow elliptical distributions; however, it cannot capture the outlyingness well when they have skewed distributions. 
To address this limitation, Dai and Genton\cite{mvod} introduced the directional outlyingness, which they also utilize in the MS plot. 

The directional outlyingness is computed by splitting the data into halves around the median and using the\,robust\,scale\,estimator to handle any skewness.
The directional outlyingness $\mathrmbf{O}$ is defined by:
\begin{equation}
    \mathrmbf{O}\left( \mathrmbf{X}(t), F_{\mathrmbf{X}(t)}\right) = \bigg\{\frac{1}{d\big(\mathrmbf{X}(t), F_{\mathrmbf{X}(t)}\big)} - 1 \bigg\} \cdot \mathrmbf{v}(t)
    \label{eq:o}
\end{equation}
where $\mathrmbf{X}$ is a $K$-dimensional function defined on a time domain $\mathscr{T}$, $F_{\mathrmbf{X}(t)}$ is a distribution of a random variable $\mathrmbf{X}(t)$, $d$ ($d > 0$) is a depth function which decides ranks of functional observations from most outlying to most typical, and $\mathrmbf{v}(t)$ is the unit vector pointing from the median of $F_{\mathrmbf{X}(t)}$ to $\mathrmbf{X}(t)$. With $\mathrmbf{Z}(t)$ as the unique median of $F_{\mathrmbf{X}(t)}$, $\mathrmbf{v}(t) = (\mathrmbf{X}(t) - \mathrmbf{Z}(t)) / \norm{\mathrmbf{X}(t)- \mathrmbf{Z}(t)}_{2}$ ($\norm{\cdot}_2$ denotes the $L^2$-norm). 
Dai and Genton~\cite{mvod} use the projection depth as a default depth function $d$. 
Note that our algorithms also employ the projection depth to follow their default.
The projection depth ($\mathrm{PD}$) is defined as:
\begin{equation}
   \mathrm{PD}\left( \mathrmbf{X}(t), F_{\mathrmbf{X}(t)} \right) = \frac{1}{1 + \mathrm{SDO}\big(\mathrmbf{X}(t), F_{\mathrmbf{X}(t)}\big)}
   \label{eg:pd}
\end{equation}
\begin{equation}
   \!\mathrm{SDO}\!\left(\mathrmbf{X}(t), F_{\mathrmbf{X}(t)} \right)\! =\!\!  \sup_{\norm{\mathbf{u}} = 1} \!\! \frac{\norm{ \mathbf{u}^{\!\top}\! \mathrmbf{X}(t) \!-\! \mathrm{median}(\mathbf{u}^{\!\top} \! \mathrmbf{X}(t)) }_2}{\mathrm{MAD}\left(\mathbf{u}^{\!\top} \! \mathrmbf{X}(t) \right)}
\end{equation}
where $\mathrm{SDO}$ is the Stahel-Donoho outlyingness~\cite{zuo2004stahel} and $\mathrm{MAD}$ is the median absolute deviation.
The three measures of directional outlyingness are defined as: 

\noindent$\boldsymbol{\cdot}$ Mean directional outlyingness ($\mathrmbf{MO}$):
        \begin{equation}
            \mathrmbf{MO}\left(\mathrmbf{X}, F_\mathrmbf{X}\right) = \int_\mathscr{T} \mathrmbf{O}\left( \mathrmbf{X}(t), F_\mathrmbf{X}(t) \right) w(t)dt
            \label{eq:mo}
        \end{equation}
        
\noindent$\boldsymbol{\cdot}$ Variation of directional outlyingness ($\mathrm{VO}$):
        \begin{equation}
            \!\mathrm{VO}\!\left(\mathrmbf{X},\! F_\mathrmbf{X}\right)\!=\!\! \int_\mathscr{T}\!\! \norm{\mathrmbf{O}\!\left( \mathrmbf{X}(t), \! F_\mathrmbf{X}(t) \right) \!-\! \mathrmbf{MO}\!\left(\mathrmbf{X},\! F_\mathrmbf{X}\right)}^2_2 \!w(t)dt
            \label{eq:vo}
        \end{equation}
\noindent$\boldsymbol{\cdot}$ Functional directional outlyingness ($\mathrm{FO}$):
        \begin{equation}
            \begin{split}
            \mathrm{FO}\left(\mathrmbf{X}, F_\mathrmbf{X}\right) & = \int_\mathscr{T} \norm{\mathrmbf{O}\left( \mathrmbf{X}(t), F_\mathrmbf{X}(t) \right)}^2_2 w(t)dt \\
            & = \norm{\mathrmbf{MO}\left(\mathrmbf{X}, F_\mathrmbf{X}\right)}_2^2 + \mathrm{VO}\left(\mathrmbf{X}, F_\mathrmbf{X}\right)
            \end{split}
            \label{eq:fo}
        \end{equation}
where $w(t)$ is a weight function defined on $\mathscr{T}$. 
Dai and Genton~\cite{mvod} use a constant weight function, i.e., $w(t) = \{\lambda(\mathscr{T})\}^{-1}$ where $\lambda(\cdot)$ represents the Lebesgue measure. 
Note that $\mathrmbf{MO}$ is simply the mean of the directional outlyingness for each of the $K$ dimensions; thus, $\mathrmbf{MO}$ is a $K$-dimensional vector.
On the other hand, $\mathrm{VO}$ and $\mathrm{FO}$ involve the computation of the $L^2$-norm, resulting in scalar values.

As shown in \autoref{fig:msplot}, we can visually identify the functional outliers with the MS plot. 
In this example, each time series (e.g., measured temperature) is from one univariate function; thus, MO is a scalar.
MO and VO show the outlyingness in magnitude and shape.
Thus, the central time series (i.e., time series similar to the other majority of time series) are mapped the region with small $|\mathrm{MO}|$ and small VO (e.g., Cluster 1 in \autoref{fig:msplot}). 
On the other hand, outliers that take different magnitudes from the others across time are located in the region with large $|\mathrm{MO}|$ and small VO (e.g., Cluster 6). 
Similarly, outliers that have a different curve shape from the others are placed in the region with small $|\mathrm{MO}|$ and large VO (e.g., Cluster 3). 
Time series with large $|\mathrm{MO}|$ and large VO (e.g., Cluster 7) are the curves with greatly outlying in both magnitude and shape.
For the case where $\mathrmbf{MO}$'s dimension is higher than 2 (i.e., $K > 2$), we can visualize the information of $\mathrmbf{MO}$ and VO with high-dimensional visualization methods, such as parallel coordinates, as Dai and Genton suggested~\cite{mvod}.
In the following, we describe our algorithm for the case where $K = 1$ as this can be considered the main visual analysis target with the MS plot. 
Note that when $K = 1$, $\mathrm{SDO}(\mathrm{X}(t), F_{\mathrmbf{X}(t)}) = | \mathrm{X}(t) - \mathrm{Z}(t) | \, / \, \mathrm{median}(|\mathrm{X}(t) - \mathrm{Z}(t)|)$.

\subsubsection{Incremental Updates of the MS Plot along Time}
\label{sec:inc_msplot}

Even though the MS plot is generated with $\mathrm{MO}$ and $\mathrm{VO}$ defined with a function, in practice, $\mathrm{MO}$ and $\mathrm{VO}$ need to be computed with a finite but large number of time points. 
As the number of time points increases, more computations are required. 
Consequently, when newly measured time points are continually fed from the data stream, the MS plot generation gradually becomes infeasible in real-time. To solve this issue, we derive equations that enable incremental updates of $\mathrm{MO}$ and $\mathrm{VO}$. 
Our incremental updates provide \textit{exact} $\mathrm{MO}$ and $\mathrm{VO}$ without any approximation. 

Using the projection depth as a depth function, the discretized versions of \autoref{eq:o}, \ref{eq:mo}, and \ref{eq:fo} are:
\begin{equation}
    \mathrm{O}\!\big( \mathrm{X}^T\![t]\big)\! =\! \mathrm{SDO}\big(\mathrm{X}^T\![t]\big) \cdot \mathrm{v}^{T}\![t]\! =\! \frac{\mathrm{X}^T[t] \!-\! \mathrm{Z}^T[t]}{\mathrm{median}(\left| \mathrm{X}^T[t] \!-\! \mathrm{Z}^T[t] \right|)}
    \label{eq:o_dis}
\end{equation}
\begin{equation}
    \mathrm{MO}^T\big(\mathrm{X}^T\big) = \sum_{t=1}^{T} \mathrm{O}\big( \mathrm{X}^T[t] \big) \, w^T[t],
    \label{eq:mo_dis}
\end{equation}
\begin{equation}
    \mathrm{FO}^T\big(\mathrm{X}^T\big) = \sum_{t=1}^{T} \mathrm{O}\big( \mathrm{X}^T[t] \big)^2 w^T[t]
    \label{eq:fo_dis}
\end{equation}
where $T$ is the number of time points available so far and superscript $T$ represents that each measure is defined on a time range $[1, T]$.
Practically, $\mathrm{Z}^T[t]$ is estimated from measured values; thus, we can assume $\mathrm{Z}^T[t]$ is simply the median of $N$ time series $\{\mathrm{X}^T_1, \cdots, \mathrm{X}^T_N\}$ at time point $t$ and $\mathrm{median}(|\mathrm{X}^T[t]-\mathrm{Z}^T[t]|)$ is the median of $\{|\mathrm{X}^T_1[t] - \mathrm{Z}^T[t]|, \cdots, |\mathrm{X}^T_N[t] - \mathrm{Z}^T[t]|\}$.
Also, here we assume $T$ time points have an approximately constant time interval.
To follow the default weight function by Dai and Genton~\cite{mvod}, we also use a constant weight function as $w^T$. 
In the discretized case, $w^T[t] = 1/T$.
    
When adding a new time point at $T+1$, \autoref{eq:mo_dis} and \autoref{eq:fo_dis} become:
\begin{equation}
    \begin{split}
        \!\!\! \mathrm{MO}^{\!T+1}\!\big(\mathrm{X}^{\!T+1}\big) 
        & \!\!=\!\! \sum_{t=1}^{T+1} \mathrm{O}\big( \mathrm{X}^{T+1}[t] \big) w^{T+1}[t] \\
        & \!\!=\!\! \frac{1}{T\!+\!1}\!\left(T \mathrm{MO}^{\! T}\big(\mathrm{X}^{\! T} \big) \!+\! \mathrm{O}\big( \mathrm{X}^{\! T+1}[T\!+\!1]\big) \right)\!,
    \end{split}
    \label{eq:mo_dis_inc}
\end{equation}
\begin{equation}
    \begin{split}
        \!\!\! \mathrm{FO}^{T+1}\!\big(\mathrm{X}^{T+1}\!\big) 
        & \!\!=\!\! \sum_{t=1}^{T+1} \mathrm{O}\big(\mathrm{X}^{T+1}[t]\big)^2 w^{T+1}[t] \\
        & \!\!=\!\! \frac{1}{T\!+\!1}\!\bigg(\! T \mathrm{FO}^{T}\!\big(\mathrm{X}^{T}\big) \!+\! \mathrm{O} \big( \mathrm{X}^{T+1}[T\!+\!1]\big)^2 \!\bigg)\!.
    \end{split}
    \label{eq:fo_dis_inc}
\end{equation}
Then, because of \autoref{eq:fo}, $\mathrm{VO}^{T+1} = \mathrm{FO}^{T+1} - (\mathrm{MO}^{T+1})^2$.

When computing $\mathrm{MO}^{T+1}$ and $\mathrm{FO}^{T+1}$, we have already calculated $\mathrm{MO}^{T}$ and $\mathrm{FO}^{T}$.
Thus, to obtain $\mathrm{MO}^{T+1}$ and $\mathrm{FO}^{T+1}$ for all $N$ time series, we need to only calculate the directional outlyingness $\mathrm{O}$ for the newly added time point.
This process has time complexity $\bigO(N)$ when using the PD as a depth function. 
Also, the required memory space to save the previous results, MO and FO,  for all $N$ time series is $\bigO(N)$. 
Therefore, we can update the exact values of the three measures of directional outlyingness with small time and space complexities that do not relate to the increasing number of time points, $T$.
Note that while the equations above are for the incremental addition, as seen in \autoref{eq:mo_dis_inc} and \autoref{eq:fo_dis_inc}, the incremental deletion is also supported, which can be derived with minor adjustments of signs, etc.

\subsubsection{Progressive Updates for New Time Series}
\label{sec:prog_msplot}

When the number of measured time series, $N$, in a system is large, the overhead of computing the directional outlyingness measures can be large. 
The original MS plot requires recomputation when new time series are added (e.g., adding temperatures obtained from different compute racks).
To provide useful intermediate results or enable the incremental addition of time series, we design a progressive algorithm that generates the MS plot with estimated directional outlyingness measures. 
We also provide a refinement mechanism to maintain the MS plot quality.

Unlike incremental updates along time (\autoref{sec:inc_msplot}), when adding new time series, we cannot obtain exact solutions while keeping the time complexity constant in terms of an increase in the number of time series.
This is because all computations are related to $\mathrm{Z}^T$ (the median of $N$ values of $\mathrm{X}^T$); thus, the update of $\mathrm{Z}^T$ based on a new time series requires recomputation of the measures for all time series at all time points.
Therefore, we (1) incrementally update the results by assuming $\mathrm{Z}^{T, N+1} \approx \mathrm{Z}^{T, N}$ (superscript $N$ shows the corresponding measure is defined on $N$ time series) as long as the errors are within the predefined error threshold and (2) progressively update the results with $\mathrm{Z}^{T, N+1}$ if the predefined error threshold is crossed.
For a newly added time series, with the assumption of $\mathrm{Z}^{T, N+1} \approx \mathrm{Z}^{T, N}$, we can easily compute $\mathrm{O}$, $\mathrm{MO}^T$, and $\mathrm{FO}^T$ with \autoref{eq:o_dis}--\ref{eq:fo_dis}. 
Also, $\mathrm{VO}^{T} = \mathrm{FO}^{T} - (\mathrm{MO}^{T})^2$.
The condition $\mathrm{Z}^{T, N+1} \approx \mathrm{Z}^{T, N}$ is checked using the Kullback–Leibler (KL) divergence of the mean absolute deviation between the new and original time series. 
Since KL divergence values can range anywhere between $0$--infinity, we have set the  error threshold $10$ by default, and this can be varied by the user. We found that an error threshold of 10 provided reasonable results without much information loss through data processing. However, the automatic identification of the error threshold is not within the scope of this work.

When adding a new time series, there is a good possibility that the time series closely follows the shape and magnitudes represented by the existing function.
By avoiding updating $\mathrm{Z}^T$ for each addition of a time series, we significantly reduce the computational overhead.
However, as stated above, once the difference between $\mathrm{Z}^{T, N}$ and $\mathrm{Z}^{T, N+1}$ becomes larger than the threshold, our algorithm starts to recompute the measures for each time series one by one. 
Similar to the incremental update in \autoref{sec:inc_msplot}, the deletion of time series is also supported with minor changes to the above equations.

\subsubsection{Implementation  with Visual and Computational Considerations on Update Frequency}
\label{sec:msplot_implemetation}

We implement our tool with the design recommendations for progressive visual analytics systems by Turkay et al.~\cite{turkay2017designing}.
In case of receiving a new time point, we update the outlyingness measures at each time point in the back-end. 
However, the MS plot is not updated at every addition of a time point; instead, we update it at a predefined number of arrived time points ($10$ by default).
When receiving a new time series, we update both the outlying measures and the MS plot. 
This procedure adds a new circle to the MS plot.
When $\mathrm{Z}^{T, N+1} \not\approx \mathrm{Z}^{T, N}$, the outlyingness measures no longer give reasonable results.
Hence, the previously computed measures need to be updated with $\mathrm{Z}^{T, N+1}$. 
This update can be computationally expensive to do on the fly if the dataset size is large. 
Thus, we compute the results asynchronously in the back-end, and the front-end visualization is updated on completion. 
All visual updates are performed with animated transitions for easy interpretation of changes.

\subsection{Reviewing the MS Plot with Auxiliary Information and Functional Principal Component Analysis (FPCA)}
\label{sec:fpca_and_visualization}

\begin{figure*}[t!]
	\centering
	\captionsetup{farskip=0pt}
    \includegraphics[width=\linewidth,height=0.57\linewidth]{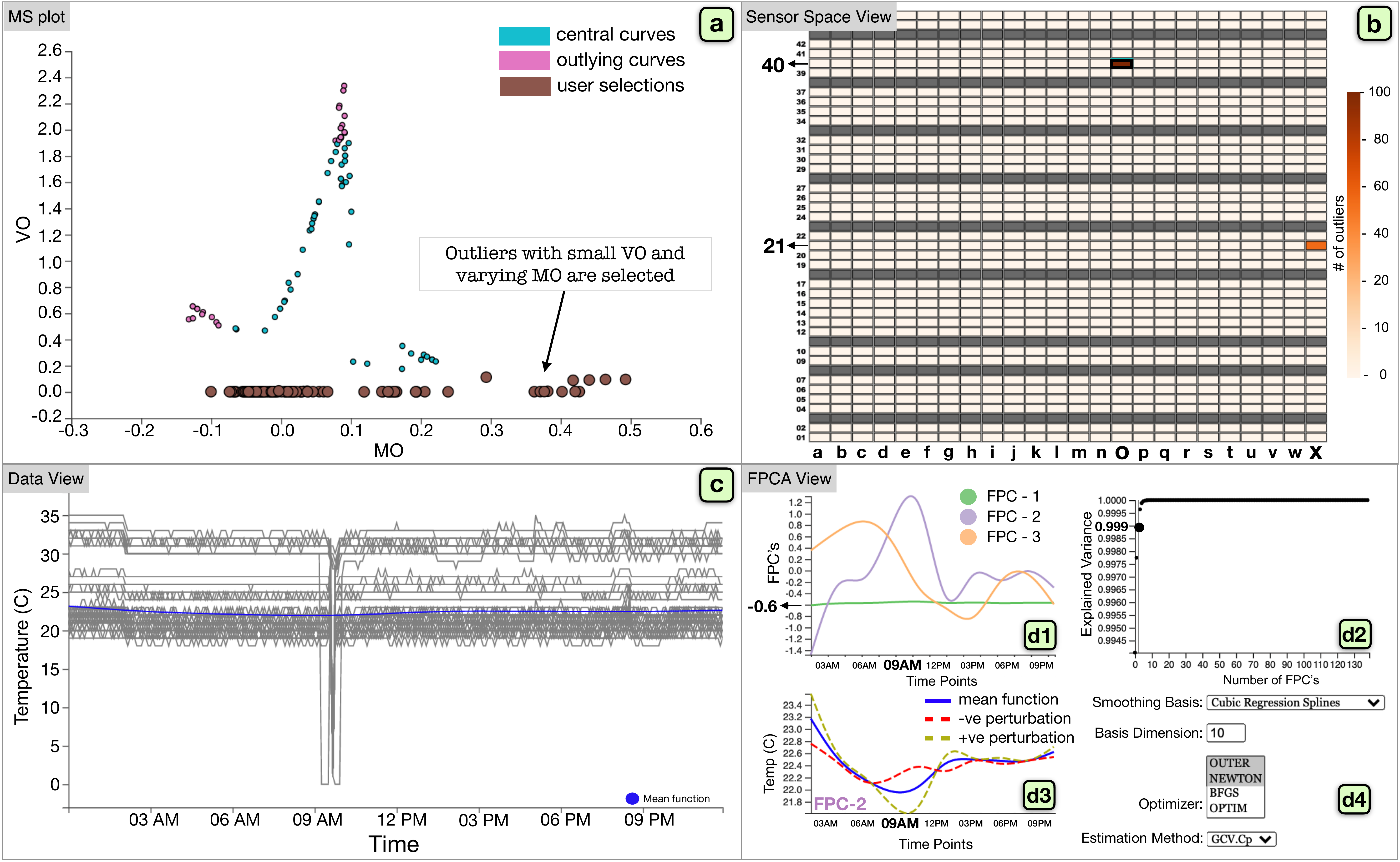}
    \caption{The UI of our visual analytics tool. (a) The MS plot shows MO and VO as a scatterplot, which is being updated incrementally and progressively. 
    (b) The sensor space view provides the related spatial information if available (here shows the compute rack information).
    (c) The data view depicts the time series selected in the MS plot.
    (d) The FPCA view displays the FPCA results and settings, which include: the (d1) FPC plot, (d2) scree plot, (d3) FPC as a perturbation of the mean plot, and (d4) functional smoothing parameter selection panel.}
	\label{fig:cs_O3}
\end{figure*}

\autoref{fig:cs_O3} shows our tool's UI, consisting of the (a) MS plot, (b) sensor space, (c) data, and (d) FPCA views. 
The analysis starts from the selection in the sensor space view, which shows the spatial information related to the data, if available.  
For example, in \autoref{fig:cs_O3}, we analyze multiple temperatures measured at each compute rack in a supercomputer---390 temperature readings per rack.
In \autoref{fig:cs_O3}-b, we visualize the locations of the racks together with 
color encoding the number of outlier readings.
With mouse actions (clicking or lasso selection), the analyst can select items of interest (i.e., racks in \autoref{fig:cs_O3}-b), and then the UI shows the related points in the MS plot. 
An example in \autoref{fig:cs_O3}-a shows the result of selecting the dark orange cell (i.e., the rack contains many outliers) around the top center (\texttt{o40}) of \autoref{fig:cs_O3}-b. Here the outlier counts are the results previously computed from the MS plot for each component.

The MS plot (\autoref{fig:cs_O3}-a) is generated with the algorithms described in \autoref{sec:inc_prog_msplot}. 
We color circles by their membership in either central or outlying curves (teal: central, pink: outlying).
For clarity, we define the central curves as curves with MO within 25-75\% of the value range and VO below 75\% of the value range. This adjustable range is selected while considering the normal operation region for sensor measurements, such as voltage and temperature.
Analysts can interactively select interesting areas (bigger brown circles) to examine more details in other views (\autoref{fig:cs_O3}-c, d).

\autoref{fig:cs_O3}-c visualizes the selections as line charts. 
As a typical central curve, a blue line shows their mean function---a smoothed line around their mean at each time point.
These visualized lines can be further analyzed with the FPCA view (\autoref{fig:cs_O3}-d). To apply FPCA~\cite{wangdavis} on the lines shown in \autoref{fig:cs_O3}-c, we first need to smooth the lines. 
\autoref{fig:cs_O3}-d4 allows analysts to select the smoothing basis functions, the number of basis, the method for optimizing the smoothing parameter, and the smoothing parameter estimation method~\cite{Higdon2013}. We show the default settings in \autoref{fig:cs_O3}-d4. 

FPCA is then performed on the smoothed lines, and FPCs are generated, as shown in \autoref{fig:cs_O3}-d1. 
Analogous to PCA, FPCs preserve the variance of functions as much as possible by defining a weight for each time point in a continuous curve.
From the shape of each FPC, we can identify the time range that has a strong influence on each FPC.
For example, in \autoref{fig:cs_O3}-d1, the first FPC (\texttt{FPC-1}) has the same weight around $-0.6$ across time, while the second FPC (\texttt{FPC-2}) has a large weight at around 9 AM.
Because most of the lines are relatively flat in \autoref{fig:cs_O3}-c, FPC-1 seems to preserve the major variance related to all the time points; FPC-2 seems to preserve the variance related to the distinct drop around 9 AM in \autoref{fig:cs_O3}-c.
This can be confirmed by our time series selection method that relates to each FPC.
When analysts select one FPC from \autoref{fig:cs_O3}-d1, we compute the FPC score~\cite{wangdavis} for each time series, which represents how strongly the time series is related to each FPC. 
Then, the tool selects and highlights the top-$k$ ($k=10$ by default) time series that highly relate to the selected FPC.
The corresponding circles are also highlighted in \autoref{fig:cs_O3}-a.
Therefore, FPCA identifies the influential time and categorizes time series using FPCs. 
While the top-3 FPCs are shown in the FPC view by default, analysts can use the scree plot (\autoref{fig:cs_O3}-d2) to select the number of FPCs to be shown. 
Similar to PCA, it shows the number of dimensions (FPCs) and the cumulative explained variance ratio. 
From this, analysts can judge how many FPCs are required to capture the data variance reasonably.

In the FPC as a perturbation of the mean plot (\autoref{fig:cs_O3}-d3), we plot the mean function of the selected data (the same one in \autoref{fig:cs_O3}-c but with a different $y$-range) and the functions obtained by adding and subtracting a suitable multiple, $\sqrt{2\xi_i}$~\cite{ramsaysilverman}, of an FPC chosen from the FPC plot, \texttt{FPC-i} ($i=\{1, 2, 3, \ldots\}$), where $\xi_i$ is the eigenvalue corresponding to \texttt{FPC}-\texttt{i}. 
The obtained functions are shown as positive and negative perturbations.
This explains the fluctuation of the measured data (i.e., $y$ values in \autoref{fig:cs_O3}-d3) the \texttt{FPC-i} captures.
For example, in \autoref{fig:cs_O3}-d3, where \texttt{FPC-2} is selected, the positive perturbation has a clear drop from the mean function around the 9 AM followed by fluctuations around the mean function, which we cannot discern in the data view.
This indicates that time series highly related to FPC-2 are mainly characterized by this variation around 9 AM, which can be confirmed with \autoref{fig:cs_O3}-c. By showing only the selected time series and their FPCA results, we can reduce the visual clutter and computational costs of FPCA.

To summarize, from \autoref{fig:cs_O3}-b, we find two racks (\texttt{o40} and \texttt{x21}) that behave abnormally and select \texttt{o40} since it includes many outliers. 
Then, from the MS plot, we select time series with low VO but varying MO. 
Indeed, from \autoref{fig:cs_O3}-c, we can see that while most time series follow the mean behavior, some show a significant drop in the temperature soon after 9 AM. 
Applying FPCA to the selected time series extracts FPCs and the modes of variation, revealing that \texttt{FPC-2} captures the variation related to the drop (\autoref{fig:cs_O3}-d1). Indeed this variation was linked to a system board failure inside (\texttt{o40}). Although this was a minor variation than the other variations in the dataset (large VO in Fig. 3-b), using FPC helped reveal this relatively smaller anomaly.
The perturbation plot of \texttt{FPC-2} (\autoref{fig:cs_O3}-d3) captures readings that flip across the mean close to 9 AM in the duration of the failure. We provide further analysis of this data in \autoref{sec:case_log}.

\section{Performance Evaluation}\label{sec:perf_evaluation}

We evaluate the performance of our algorithms for two example use scenarios.
In each scenario, we mimic a practical streaming analysis situation by feeding time points and time series from a real-world dataset.
For the experiments, we use the MacBook Pro (13-inch, 2019) with a 2.8 GHz Quad-Core Intel Core i7 processor and 16 GB 2133 MHz LPDDR3 memory.
Completion times are averaged over $10$ executions.

\noindent\textbf{Evaluation with supercomputer hardware logs.}
Supercomputer hardware logs, which contain readings of voltages, temperatures (water/air/CPU), fan speeds, are collected from various sensors housed in the compute rack subsystem (e.g., system board and air/water cooling). 
Studying these logs to identify failure patterns can aid in improving the robustness and reliability of large-scale machines.
There have been past efforts to analyze supercomputer/cloud hardware logs with visual analytics tools~\cite{guo2018valse,shilpika2019mela,tfujiwara2020,cloudet,ensemblelens}. 
Our logs (analyzed in \autoref{sec:methodology}) are collected from the K computer~\cite{miyazaki2012overview} and contain data from 864 compute racks, with 1,163 different sensor readings per rack collected every 5-minute interval (i.e., 288 timestamps in a day).

We examine our incremental and progressive algorithms' efficiency for this log data, which has extremely high volume but relatively low velocity (5-minute interval). 
We assume that we have already had the data of size $4,032\!\times\!336,960$ (i.e., $2$ weeks and $336,960$ temperature readings across all racks).
Then, we add a newly arrived time point to this existing data. 
While the update of the MS plot without our incremental algorithm (i.e., recalculation on $4,033$ time points) is finished in $8.2$ minutes, the incremental addition of a time point is completed in $2.5$ seconds.  
To evaluate our progressive update with approximation, we first process the same data, but one time series is excluded (i.e., the size of $4,032\!\times\!336,959$). 
Then, we add one time series and update the MS plot using our algorithm with approximation, which is completed only in $16.6$ seconds, while the overall update without approximation is done in $8.4$ minutes. 
For both cases, the updates without our algorithms exceeded the data collection interval and caused wastage of computing resources. 

This proof-of-concept experiment shows that our algorithms achieve prompt updates for large datasets. 
In practice, the analysts may select a smaller dataset (e.g., 1 day instead of 2 weeks), and the ordinary MS plot may update the result within the current update interval (i.e., every 5 minutes). 
However, most recent supercomputers collect sensor data at $0.03$--$10$Hz~\cite{fugaku,ANL} (i.e., about every 0.1--30 seconds).
Our approach can process large volumes of data to promptly identify anomalies in these large-scale systems.

\vspace{5pt}
\noindent\textbf{Evaluation with Biometrics data.}
The generic design of our tool enables us to monitor time series collected from different types of hardware systems. 
As another use scenario, we analyze data collected from smartwatch accelerometer sensors of $51$ subjects~\cite{biometrics}. 
The subjects performed daily activities, including ambulation (e.g., walking, dribbling), fast hand-based (e.g., brushing), and eating activities.
Each subject contributed $54$ minutes of data collected at $20$Hz sensor polling rate for a total of $3$ minutes per activity.
Each row in the dataset contains \textit{Subject-id, Activity Code, Timestamp, x, y, z}. 
For brevity, we use the sensor value for the \textit{x}-axis, i.e., signed acceleration along the \textit{x}-direction. However, one can also individually apply our algorithms to each of the other measurements of \textit{x}, \textit{y}, and \textit{z}. 

Here we assume that we have already collected the data of size $30,000\!\times\!5,000$ (i.e., 25 minutes and 5,000 subject-activity pairs).
Then, we incrementally add $1,200$ time points (i.e., every $1$ minute).
While the incremental additions of $1,200$ time points are completed in $0.8$ seconds, the update without using our incremental algorithm is finished in $18$ seconds.
For our progressive update with approximation, we further add $10$ time series into this data, and the update is done in $5.5$ seconds. Without the progressive update, the overall process time was $19$ seconds. 
These results show that updates with our algorithms are fast enough to handle a large number of time series with extremely high velocity (i.e., $20$Hz, or $0.05$-second interval) in real-time.

\vspace{5pt}
\noindent\textbf{Experimental evaluation with different data sizes.}
We further evaluate our algorithms with different numbers of time points and time series while using the supercomputer hardware logs (\texttt{SC Log}) and the biometrics data (\texttt{Biometrics}).
Tables \ref{table:cINCtime} and \ref{table:cPRGtime} show  the completion times for the addition of new time points and new time series, respectively. 

In \autoref{table:cINCtime}, we first process an initial set of data consisting of varying time points (100--20,000) and a fixed number (1,000) of time series. 
We then add one new time point and update the existing outlyingness measures incrementally.
The completion time for the initial data fit increases as time points are increased, while the incremental addition has the completion time \textit{independent} of the data size.
The incremental addition always takes about $1$ms for \texttt{SC Log} and $180$ms for \texttt{Biometrics}, which are fast enough to support online streaming analysis. 
Note that \texttt{Biometrics} is characterized by widely varying high-frequency fluctuations, and the data complexity contributes to the computation time where the bottleneck relates to the depth measures. 

\begin{table}[tb]
\scriptsize
\centering
\caption{Completion time (in seconds) of the initial data fit (Initial Fit) and the incremental addition of one time point (Partial Fit).}
\label{table:cINCtime}
\footnotesize
\begin{tabular}[b]{lllll}
\hline
Dataset & $N$ & $T$ &  Initial Fit & Partial Fit \\
\hline
\texttt{SC Log} & 1,000 & 100 & 0.0109 & 0.0009 \\
\texttt{SC Log} & 1,000 & 1,000 & 0.0640 & 0.0011 \\
\texttt{SC Log} & 1,000 & 10,000 & 0.5373 & 0.0013 \\
\texttt{SC Log} & 1,000 & 20,000 & 1.3357 & 0.0010 \\
\texttt{Biometrics} & 1,000 & 100 & 0.0457 & 0.182 \\
\texttt{Biometrics} & 1,000 & 1,000 & 0.388 & 0.173 \\
\texttt{Biometrics} & 1,000 & 10,000 & 2.537 & 0.196 \\
\texttt{Biometrics} & 1,000 & 20,000 & 4.990 & 0.176 \\
\hline
\end{tabular}
\end{table}

\begin{table}[tb]
\tiny
\centering
\caption{Completion time (in seconds) of the initial fit and progressive addition of one time series with and without approximation.}
\label{table:cPRGtime}
\footnotesize
\setlength\tabcolsep{4.0pt}
\begin{tabular}[b]{llllll}
\hline
Dataset & $N$ & $T$ & Initial Fit & Approx. & Non-approx. \\
\hline
\texttt{SC Log} & 100 & 1,000 & 0.0038 & 0.0006 & 0.0088 \\
\texttt{SC Log} & 1,000 & 1,000 & 0.0495 & 0.0039 & 0.0654\\
\texttt{SC Log} & 10,000 & 1,000 & 0.5910 & 0.0680 & 1.0079\\
\texttt{SC Log} & 20,000 & 1,000 & 1.7038 & 0.1009 & 2.5148\\
\texttt{Biometrics} & 1,000 & 1,000 & 0.0520 & 0.0027 & 0.1890\\
\texttt{Biometrics} & 5,000 & 1,000 & 0.3350 & 0.0184 & 1.1533\\
\texttt{Biometrics} & 10,000 & 1,000 & 0.7328 & 0.0217 & 1.690\\
\hline
\end{tabular}
\end{table}

For the time series addition, as shown in \autoref{table:cPRGtime}, we first process an initial set of data consisting of varying time series (100--10,000) and a fixed number of time points (1,000). 
We then add one new time series and update the existing outlyingness measures using the progressive algorithm with and without the approximation.

With the approximation, it takes less than about $100$ms to compute.
Without the approximation (i.e., equivalent to an ordinary plot), the overall computation time would be almost close to the initial fit of the data, leading to more significant wait times.

\section{Use Scenarios}\label{sec:case_studies}

We have shown an analysis example in \autoref{sec:fpca_and_visualization}.
We further demonstrate our tool’s effectiveness by performing analyses for the two use scenarios described in \autoref{sec:perf_evaluation}. A demonstration video of the interface is available at~\cite{supp}.

\begin{figure}[t]
	\centering
    \includegraphics[width=\linewidth]{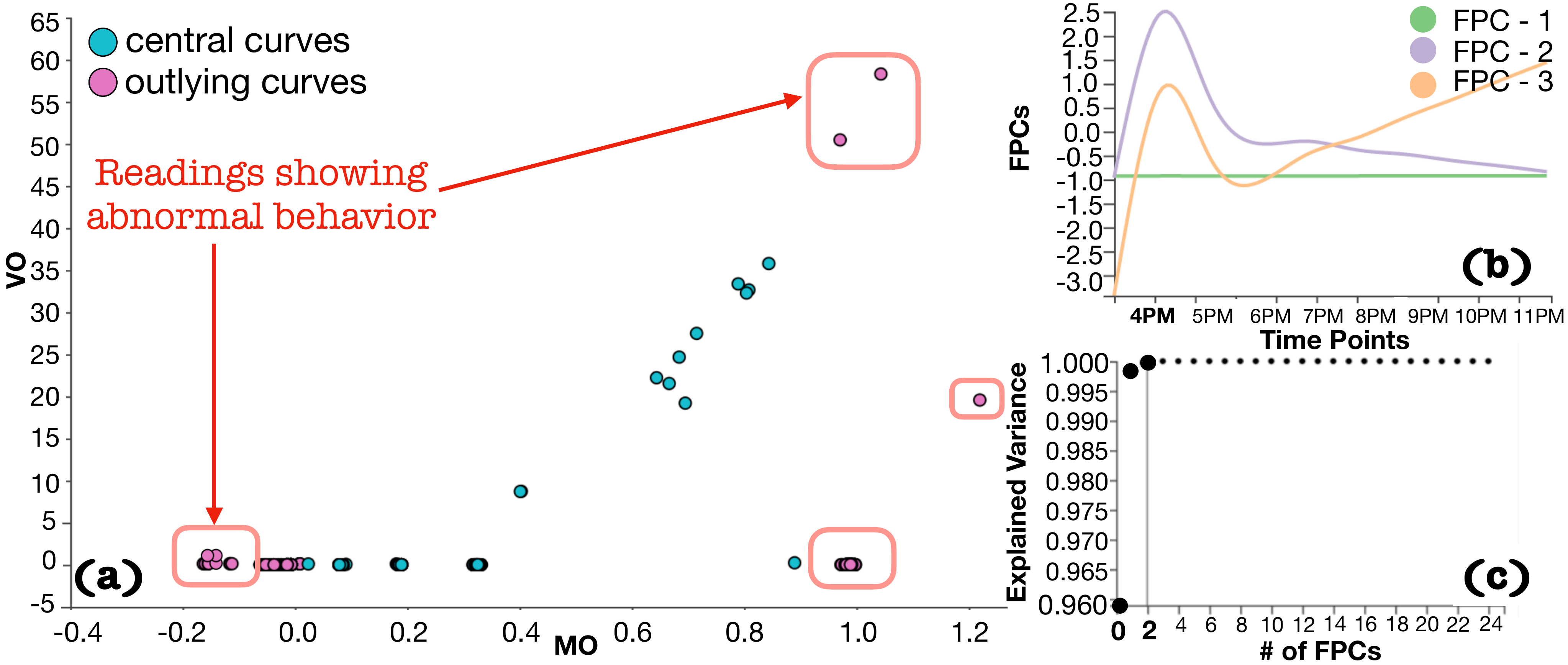}
    \caption{Analyzing the supercomputer hardware log data with the (a) MS, (b) FPC, and (c) scree plots. The central and the outlying curves represent readings with normal and abnormal behaviors, respectively.}
	\label{fig:cs_O4_1}
\end{figure}

\begin{figure*}[b]
	\centering
    \includegraphics[width=\linewidth]{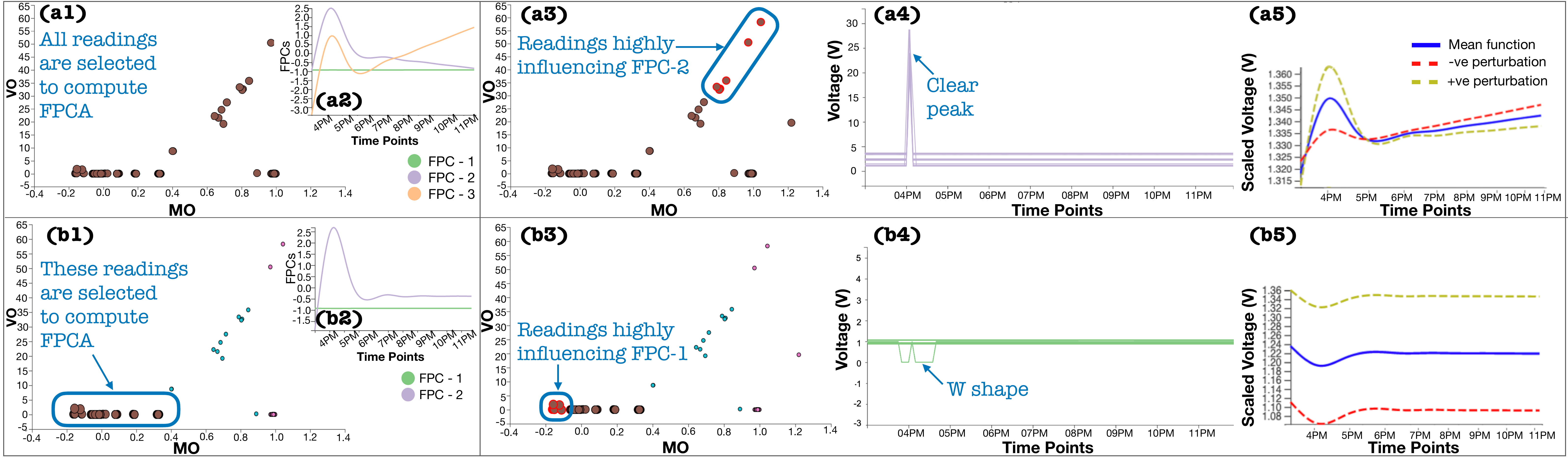}
    \caption{All readings are selected in the MS plot (a1) and then three FPCs are computed with the selected readings (a2). From FPCs, \texttt{FPC-2} is selected (a2). The corresponding readings are highlighted in the MS plot (a3) and visualized in the data view (a4). The FPC as a perturbation of the mean plot is updated accordingly (a5).
    The similar procedure is applied to (b1--b5), where two FPCs are generated with the readings of low MO and low VO (b1, b2) and then \texttt{FPC-1} is selected (b3--b5).}
	\label{fig:cs_O4_2}
\end{figure*}

\subsection{Outlier Identification and Validation from Supercomputer Hardware Logs}
\label{sec:case_log}

From the supercomputer hardware logs, we mainly want to analyze temperature and voltage readings.
There are $390$ temperature and $480$ voltage readings per rack.
Because these have different numbers of readings and are recorded by different sensors, we handle a collection of the temperature and voltage readings as a univariate function (e.g., voltage: $N\!=\!480\!\times\!864$ and $K\!=\!1$, where $864$ is the number of racks).
As we have already analyzed temperature readings in \autoref{sec:fpca_and_visualization}, here we analyze voltage readings.

\autoref{fig:cs_O4_1}-a shows the MS plot for $390$ voltage readings selected from three racks, \texttt{l07}, \texttt{l14}, and \texttt{l19}, from the space view based on our domain knowledge and interest.
Each circle in the MS plot corresponds to one time series of readings. 
The central curves (shown in cyan) represent readings that show normal behavior.
\autoref{fig:cs_O4_1}-a is the result after starting to feed new time points of the selected racks at the beginning of the day (e.g., 12 AM).
With the incremental updates, the readings showing abnormal behavior (shown in the pink box) are gradually pushed towards the outer regions of the MS plot.
\autoref{fig:cs_O4_1}-b and c show the FPCA view for all the $390$ voltage readings selected from the MS plot.
From \autoref{fig:cs_O4_1}-c, we see that the first three FPCs explain almost $100\%$ of the variation in the dataset. \autoref{fig:cs_O4_1}-b shows the three FPCs. 
\texttt{FPC-1} captures $96\%$ of the variation and is mostly flat, indicating that a larger number of the readings behave as expected without much variation.
\texttt{FPC-2} and \texttt{FPC-3} show a spike from 4 PM--4:30 PM, increasing to a magnitude of $2.5$ for \texttt{FPC-2}. \autoref{fig:cs_O4_1}-c shows that \texttt{FPC-2} represents roughly $4\%$ of the variation. 
We can expect that a smaller number of readings accounting for $4\%$ of the total variation behave peculiarly around the time 4 PM--4:30 PM.

In \autoref{fig:cs_O4_2}, we select \texttt{FPC-2} and \texttt{FPC-1}, and the linked views are updated accordingly, \autoref{fig:cs_O4_2}-a for (\texttt{FPC-2}) and \autoref{fig:cs_O4_2}-b for (\texttt{FPC-1}). \autoref{fig:cs_O4_2}-(a1,b1) shows user selections and \autoref{fig:cs_O4_2}-(a2,b2) are the corresponding FPCs.
To review more details of the user-selected readings in  \autoref{fig:cs_O4_2}-a1, related to \texttt{FPC-2}, we select \texttt{FPC-2} from  \autoref{fig:cs_O4_2}-a2; then, the readings that heavily influence \texttt{FPC-2} are shown in \autoref{fig:cs_O4_2}-(a3,a4). In \autoref{fig:cs_O4_2}-a3, the corresponding readings are highlighted with red outlines, and they have large MO and VO. In \autoref{fig:cs_O4_2}-a4, the readings show a spike at around 4 PM, lasting for about $30$ minutes. 
\autoref{fig:cs_O4_2}-a5 shows how these readings fluctuate with respect to the overall mean of the selected readings (shown in navy blue). 
This plot helps us identify the overall trend of the readings when compared to the mean behavior of the selected readings. 
The readings show an overall flipping across the mean function at 4 PM.
This example demonstrates how the MS plot captures outliers and how the FPCA view can also find similar outliers to validate the findings from the MS plot.

In \autoref{fig:cs_O4_2}-b1, we select the area with low MO and low VO. \autoref{fig:cs_O4_2}-b2 shows $2$ FPCs that capture all the data variation.
As this area is expected to follow the main trend because of low VO, we choose \texttt{FPC-1} and visualize the top-$5$ readings that heavily influence \texttt{FPC-1} (\autoref{fig:cs_O4_2}-b3, b4), which contain the ``W'' shape along with flat curves.
This variation of the curves shown in \autoref{fig:cs_O4_2}-b4 is smaller when compared to the peak in \autoref{fig:cs_O4_2}-a4.
The MS plot captures this irregular pattern by placing these readings in the negative MO and slightly higher VO, as highlighted in \autoref{fig:cs_O4_2}-b3 with red outlines. 
In \autoref{fig:cs_O4_2}-b5, the FPC perturbation to mean plot captures this small variation and the overall trend with respect to the mean behavior.
Using this example, we show that incremental MS plot and FPCA help identify intra-cluster trends by grouping similar trends within the same spatial locality of a cluster.

\subsection{Monitoring and Categorizing Biometrics of Daily Activities with Outlyingness}
\label{sec:case_biometrics}

\begin{figure}[tb]
	\centering
    \includegraphics[width=1\linewidth]{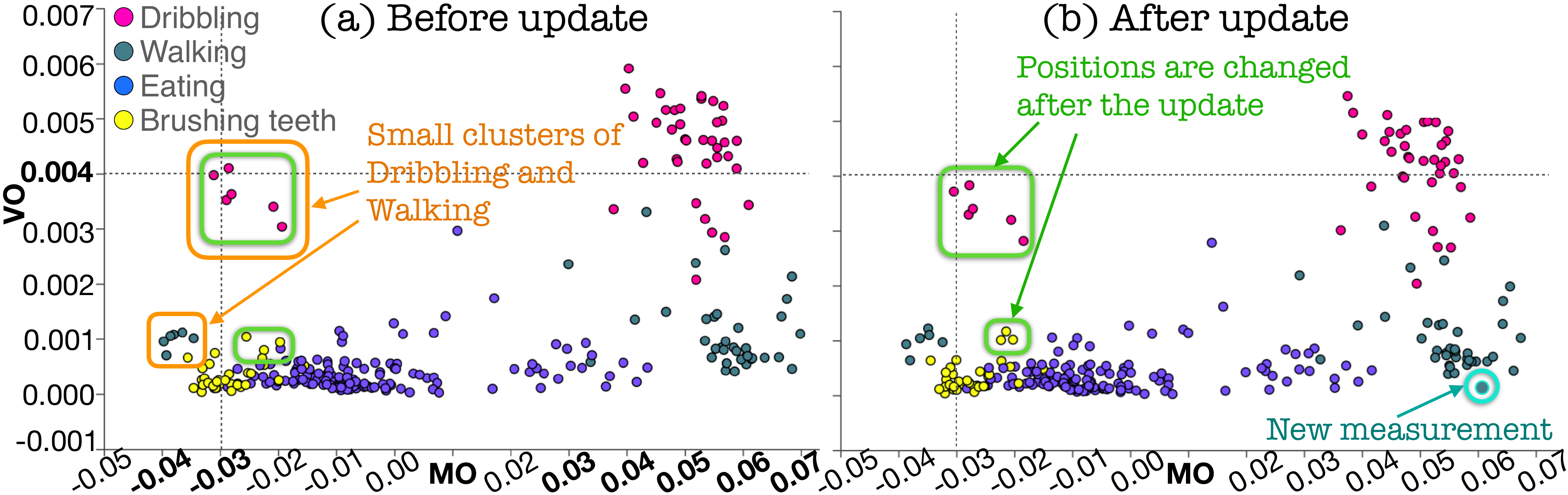}
    \caption{The MS plot for accelerometer measurements (a) before and (b) after the update. 
    (a) is generated with $1,000$ time points ($50$ seconds), and (b) with additional $100$ time points and 1 measurement.
    }
	\label{fig:inc_update}
\end{figure}

\begin{figure}[tb]
	\centering
    \includegraphics[width=1\linewidth]{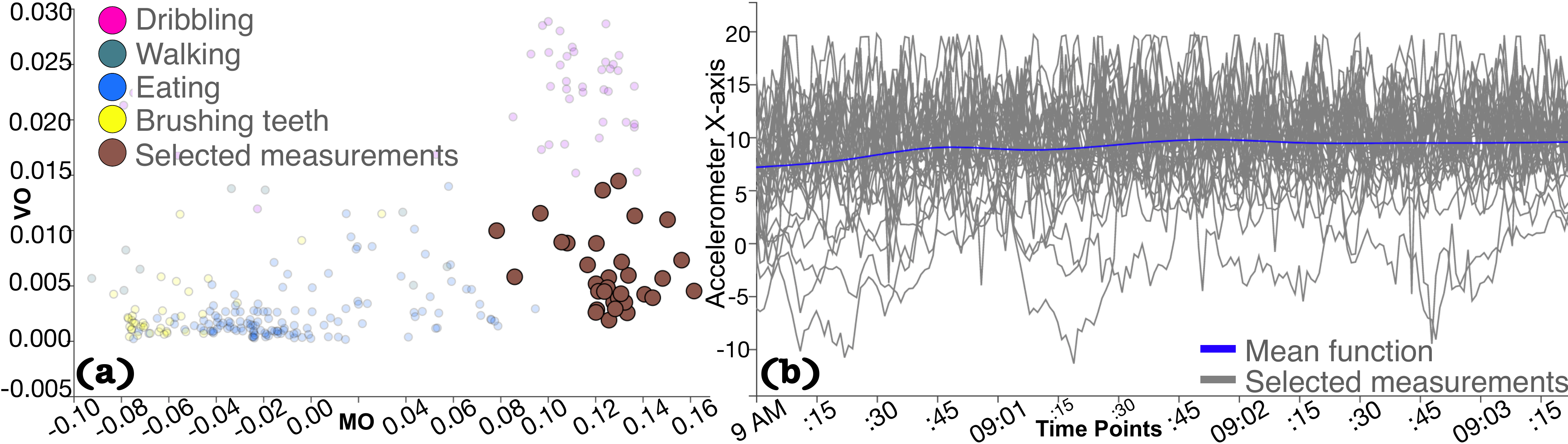}
    \caption{The (a) MS plot and the (b) data view. The data view is shown for selected measurements (brown) from (a).}
	\label{fig:sub_group}
\end{figure}

\begin{figure*}[b]
	\centering
    \includegraphics[width=1\linewidth]{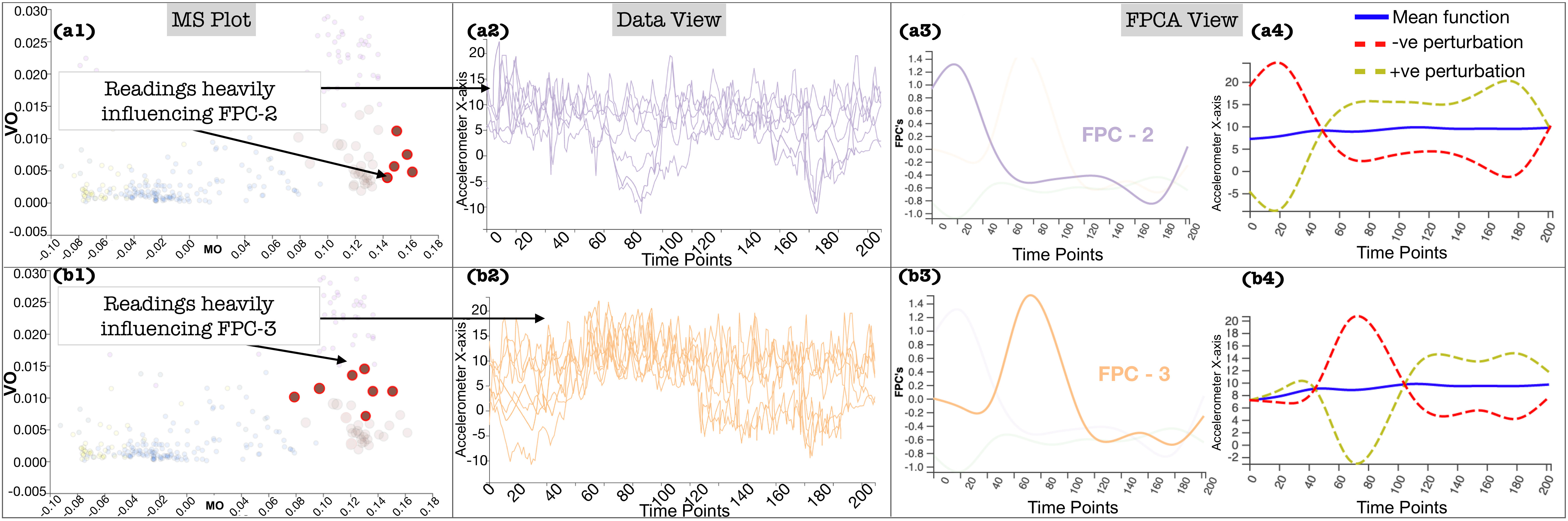}
    \caption{The MS plot, data view, FPCA view after the selections of \texttt{FPC-2} (a1--a4) and \texttt{FPC-3} (b1--b4) from \autoref{fig:sub_group}}.
	\label{fig:fpca-bio}
\end{figure*}

\autoref{fig:inc_update}-a shows the MS plot generated from $1,000$ time points (50 seconds) for subjects performing activities, such as dribbling, walking, eating, and brushing teeth. 
Here the circles in the MS plot are colored based on the subjects' activities.
We see that the MS plot generally groups the different activities well. 
This result is reasonable because each activity involves different frequencies and magnitude of changes in the subject's hand position.
For example, when compared with brushing teeth and eating, dribbling and walking involve much larger movements, resulting in high MO.  
Also, as dribbling is the only sports activity, we can expect that it has a significantly different shape of time series from the other activities; thus, dribbling has high VO.

However, we find that the individual walking and dribbling activities are separated into two clusters. 
One cluster in each activity is placed at the far left of the MS plot compared to the other, as shown in orange box. 
On examining MO values, we notice that each cluster placed at the left side has negative MO values but similar magnitudes of MO with the corresponding other cluster (e.g., for walking, one has MO values from $-0.04$ to $-0.03$ while another has values from $0.03$ to $0.07$).
Also, these two clusters for each activity have similar VO values.
From these observations, we can consider that the subjects in those clusters with negative MO values might have worn the smartwatch differently from most subjects; consequently, $x$ measurements recorded the opposite signs from the others. 
This demonstrates the usefulness of the MS plot to find these anomalies visually.

While performing incremental updates, we observe that circles corresponding to either dribbling or brushing teeth show fluctuations in the MS plot. 
For example, in \autoref{fig:inc_update}-b, where the MS plot in \autoref{fig:inc_update}-a is updated with $100$ additional time points, we see that the circles shown in green box have small but noticeable changes in their positions (e.g., the annotated \texttt{Dribbling} cluster is moved down).
This is likely because dribbling and brushing activities involve frequent hand movements when compared to the other activities.
The collective use of the MS plot and the animated transitions (described in \autoref{sec:msplot_implemetation}) helps us capture these small changes that occurred across the updates.
\autoref{fig:inc_update}-b also shows a new measurement (annotated by teal circle) added to the MS plot using the progressive update for a subject performing the walking activity. 
Although the new addition is made with the approximation, it is reasonably close to the cluster for walking activities.
Further evaluations of the effect of the approximation are presented in \autoref{sec:disc}.

Next, we review a cluster using the MS plot and FPCA together.
As an example, as shown in \autoref{fig:sub_group}-a, we select a cluster (highlighted by brown) corresponding to walking with large MO and relatively small VO.
For the sake of clarity, we have chosen to show the MS plot after obtaining the first $200$ time points ($10$ seconds). 
\autoref{fig:sub_group}-b shows the data view for the selected time series/circles. 
Because of the visual clutter caused by many curves, not much can be deduced from the current data view. 
Using FPCA, we can still identify trends and patterns in such cluttered time series.
We apply FPCA with the default setting and observe that the first 3 FPCs correspond to $90\%$ of the variation in the selected time series.
We select \texttt{FPC-2} and \texttt{FPC-3}, and the linked views are updated accordingly, as shown in \autoref{fig:fpca-bio}-a (\texttt{FPC-2}) and \autoref{fig:fpca-bio}-b (\texttt{FPC-3}). \autoref{fig:fpca-bio}-(a1,b1) and \autoref{fig:fpca-bio}-(a2,b2) show the MS plot views and the data view for the biometrics data, respectively.
\texttt{FPC-2} (\autoref{fig:fpca-bio}-a3) has a distinct peak at earlier time points followed by a drop at later time points, while
\texttt{FPC-3} (\autoref{fig:fpca-bio}-b3) has a distinct peak at around time point 80.
Similar patterns can also be seen in \autoref{fig:fpca-bio}-a4 and b4, which inform the positive and negative perturbations of the measurements with respect to the mean function of the selected subgroup. 
These peaks clearly capture a phase shift in the subjects' walking patterns. 
The corresponding MS plot and data views (\autoref{fig:fpca-bio}-a1, a2, b1, and b2) show the measurements that heavily influence the selected FPCs.
Now, the phase shift is visible in the data view. 
For example, in \autoref{fig:fpca-bio}-b2, we can see that multiple subjects tend to have larger values around time point 80 than at other time periods.
Also, from the MS plot view in \autoref{fig:fpca-bio}-b1, we can see that the patterns in \texttt{FPC-3} tend to be seen only in the circles with larger VO than the other selected circles. 
Similarly, the circles that heavily influence \texttt{FPC-2} can be seen only in the area with large MO.

To summarize, a combination of the incremental, progressive MS plot and FPCA helps filter distinct patterns in time series data of activities. 
Certain activities form a unique cluster of patterns or an anomaly cluster.
Analyzing these clusters further show us what these patterns look like and where they lie within the corresponding cluster. 
While this scenario uses the measurements from daily activities, similar high-velocity data is often collected, for example, for clinical care~\cite{fong2018recovering}, and our approach would be useful to monitor the measurements to make time-critical decisions.

\section{Discussion}
\label{sec:disc}

We have presented our approach's effectiveness to analyze streaming time series data and its applicability to various datasets and future use. We further discuss the strengths of the MS plot by comparing it with DR methods often used to categorize time series data visually~\cite{ali2019clustering}.
We provide expert feedback to support the effectiveness of our tool. 

\begin{figure*}[tb]
	\centering
    \includegraphics[width=1\linewidth]{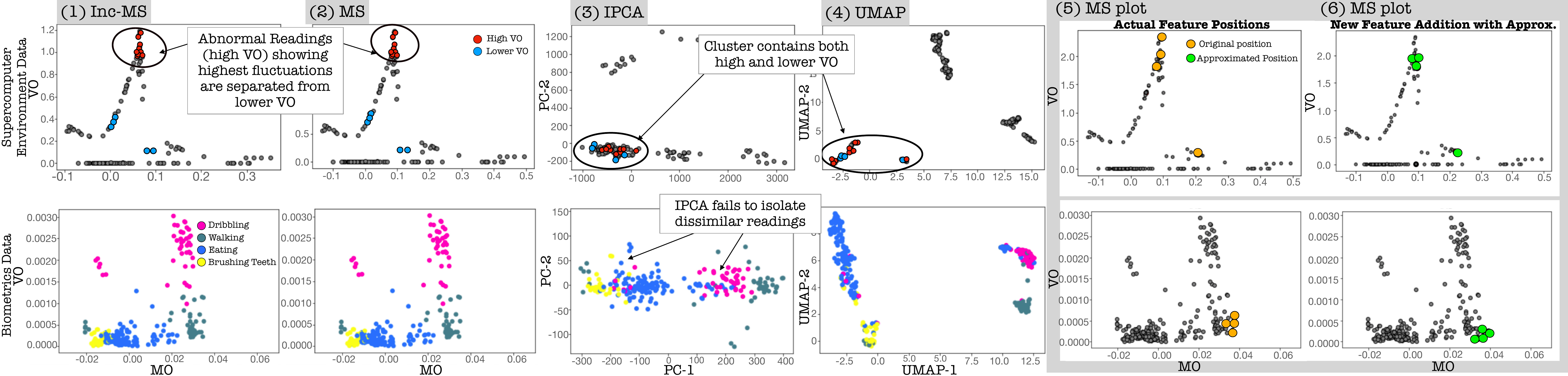}
    \caption{The comparison of results among the (1) incremental MS plot (\texttt{Inc-MS}), (2) MS plot (\texttt{MS}), (3) incremental PCA (\texttt{IPCA}), and (4) \texttt{UMAP}. (5) shows the exact position of features obtained by \texttt{MS} and (6) shows the progressive addition of the same features with approximation.}
	\label{fig:comp}
\end{figure*}

\vspace{5pt}
\noindent\textbf{Comparison with DR methods.}
\autoref{fig:comp} compares 2D plots generated by using
ordinary MS plot~\cite{mvod} (\texttt{MS}), our incremental MS plot (\texttt{Inc-MS}), incremental PCA (\texttt{IPCA})~\cite{ross2008incremental}, and \texttt{UMAP}~\cite{mcinnes2018umap} for supercomputer hardware logs and biometrics data to show how these methods capture the underlying data patterns. \texttt{IPCA} and \texttt{UMAP} are chosen as representative linear and nonlinear DR methods that are used for visual pattern identification in multiple time series~\cite{ali2019clustering,tfujiwara2020}.

First, as expected, for both datasets, \texttt{MS} and \texttt{Inc-MS} produce identical results since \texttt{Inc-MS} does not apply any approximation.
For the supercomputer logs, we compare the four methods' ability to discern the outlier readings with large VO (abnormal) and lower VO (normal). To avoid visual clutter, we ignore the rest of the readings shown in gray color. \texttt{UMAP} and \texttt{IPCA} produce discernible clusters. 
Although most of the clusters group readings with similar trends, some of the clusters show both normal (blue) and abnormal (red) readings (verified by using the data view in our tool). 
This issue can occur as they are not specifically designed to capture outlyingness.
For the biometrics data, while \texttt{MS} and \texttt{UMAP} isolate different activities into clusters, \texttt{IPCA} fails to do so. 
The \texttt{UMAP} result with the setting (specifically, $\mathrm{n\_neighbors}\!=\!4$, $\mathrm{min\_dist}\!=\!0.1$, and $\mathrm{metric}$=``euclidean'') that showed reasonable clarity between clusters still show larger overlap in the measurements from different activities (e.g., dribbling and eating) when compared to \texttt{MS}.
Another strength of the MS plot is in its interpretability.
Since the MS plot directly shows the magnitude and shape outlyingness in the $x$- and $y$-coordinates, we can easily interpret why some time series are grouped together (as demonstrated in \autoref{sec:case_studies}). 
On the other hand, DR methods such as incremental PCA and UMAP require additional steps to understand each revealed cluster~\cite{fujiwara2020supporting}. 
In \autoref{fig:comp}-5 and 6, we examine the effect of approximation in our progressive algorithm by comparing it to the exact solution obtained with \texttt{MS}.
The added time series are colored orange and green in each result.
When the errors are within the user-specified error threshold, the approximations can be assumed to produce close to valid results. 
In fact, as shown in \autoref{fig:comp}-5 and 6, the newly added time series are in close vicinity of the actual values. 

\begin{figure}[t]
	\centering
    \includegraphics[width=0.95\linewidth]{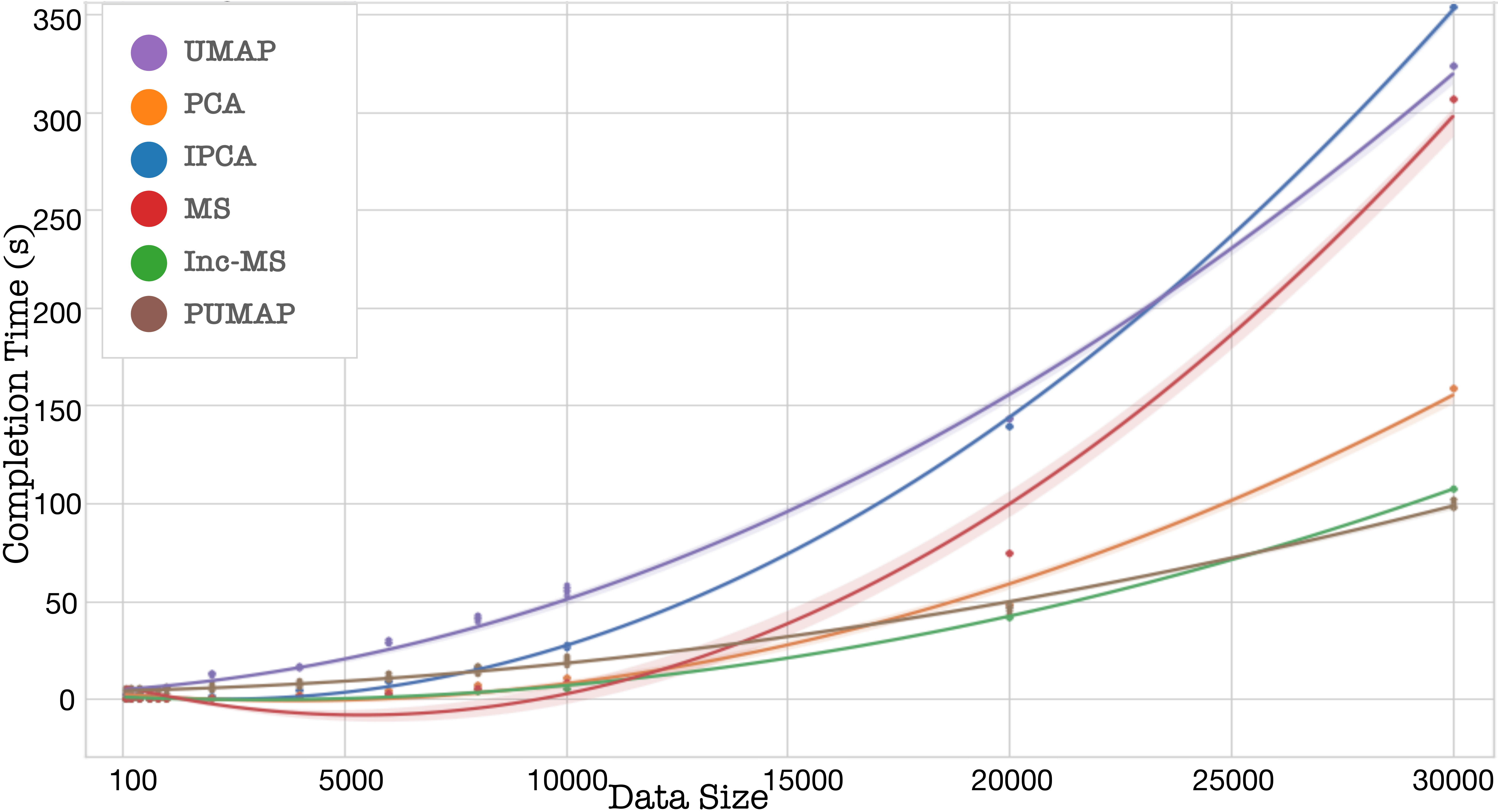}
    \caption{The comparison of completion time, showing how performance scales with datasize.}
	\label{fig:perf}
\end{figure}

In \autoref{fig:perf}, we use the supercomputer hardware logs to compare the completion time of methods and their streaming counterparts used in \autoref{fig:comp}. 
For this evaluation, we use the same experimental platform as the one used in \autoref{sec:perf_evaluation}.
We scale the data starting with a data size of $N\!\times\!T \!=\!100\!\times\!100$ to $30,000\!\times\!30,000$.
We use scikit-learn's implementation for \texttt{IPCA}, McInnes et al.’s implementation for \texttt{UMAP}~\cite{mcinnes2018umap}, and Ko et al.'s implementation for \texttt{PUMAP}~\cite{pumap}.
\texttt{Inc-MS} provides a method that incrementally updates the results, which is called \textit{partial fit}, and a method that processes input data all at once called \textit{(initial) fit}. We start by processing $5\%$ of the time points with the \textit{initial fit} and use \textit{partial fit} containing $5$ time points for the rest.
We use a regression plot to interpolate the results.
In~\autoref{fig:perf}, for each data size, the \texttt{Inc-MS} plot either gave approximately the same performance as \texttt{MS} (for data size in the range of hundreds) or outperformed \texttt{MS} by a factor of $2$ times for a data size of $20,000\times20,000$ and  $3$ times for the data size of $30,000\times30,000$. We have improved the original MS algorithm by updating the results already computed rather than recomputing the overall result every time a new time point is added; this significantly improves performance. \texttt{Inc-MS} implementation outperforms \texttt{MS}, \texttt{PCA}, \texttt{IPCA}, and \texttt{UMAP} while having a runtime close to \texttt{PUMAP}. \texttt{PUMAP} provides approximated results, but \texttt{Inc-MS} provides exact results. Note that as $N$ increases, \texttt{PCA} outperforms \texttt{IPCA}~\cite{tak2019}. 

\vspace{5pt}
\noindent\textbf{Expert feedback.}
We evaluated our tool with two industry experts with over 5 years of experience analyzing large-scale computing systems' hardware logs daily.
We installed the tool locally on their system and provided the details of each component's functionalities. 
The experts were impressed by our overall design and provided a detailed assessment of its components. 
They had no trouble interpreting the details presented in each view. 
For the overall system, they commented:
``{\it The updates to the MS plots showed a smooth transition and were easier to follow. 
Upon selecting a cluster of measurements in the MS plot, we can view where the similar time series lie on the system in the space view and what pattern they represent in the data view. 
The FPC plots provide a summarized representation of the selected data, which was helpful.}''
They indicated that the UI provides a good intuition of the overall system behavior in real-time. 
The overall feedback is encouraging, although they also noticed some limitations of the tool.
Upon selecting from the FPC plot, the measurements that highly influence the selected FPC are highlighted in the data view and the MS plot. 
However, the experts mentioned that it would be useful to view components having little to no influence on this FPC as well. 
At the time of writing this paper, we have added a range slider that controls the threshold of selecting influential time series (circles, curves) on FPCs.

\section{Conclusion}

The advent of exascale systems makes it pertinent to build applications that can process data in a timely and reactive manner. 
We have built a visual analytics tool that processes large streaming time series data, which are intrinsically functional, using functional data analysis. 
This is achieved through incremental and progressive computations of the magnitude-shape outlyingness measure through the addition of new time points or new time series.
To further understand the underlying errors, we use functional principal component analysis to identify different modes of variation and the type of fluctuation (amplitude, phase shift) for each mode. 
We plan to extend our tool's visualizations to support the analysis of multivariate functions and also incremental and progressive algorithms to use different depth measures.



\ifCLASSOPTIONcompsoc
  \section*{Acknowledgments}
\else
  \section*{Acknowledgment}
\fi


This research was supported in part by the U.S. National Science Foundation through grant IIS-1741536 and the Argonne National Laboratory through contract 8F-30225. We thank Keiji Yamamoto for providing useful information regarding the supercomputer data. 

\bibliographystyle{IEEEtran}
\bibliography{template}

\vspace{-0.4in}
\begin{IEEEbiography}[{\includegraphics[width=1in,height=1.25in,clip,keepaspectratio]{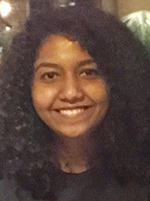}}]%
    {Shilpika} received the MS degree in Computer Science from Loyola University of Chicago. She is a Ph.D. student in computer science at the University of California, Davis. Her main research interests include in the fields of data science and data visualization. Her research focuses on using machine learning for anomaly detection and prediction in high performance computing systems.
\end{IEEEbiography}
\vspace{-0.35in}

\begin{IEEEbiography}[{\includegraphics[width=1in,height=1.25in,clip,keepaspectratio]{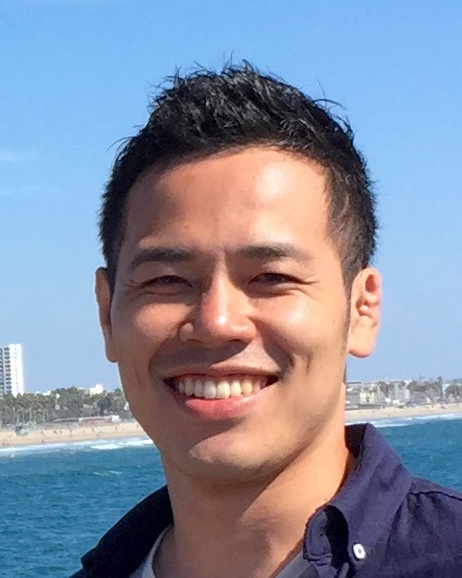}}]%
    {Takanori Fujiwara} received his Ph.D. degree in computer science at the University of California, Davis in 2021. Before UC Davis, he received a Master's degree in environmental studies in 2011 from the University of Tokyo and worked for Kajima Corporation in Japan. He works at the intersection of data science and visualization with special interests in visual analytics using ML for high-dimensional and network data. 
\end{IEEEbiography}
\vspace{-0.35in}

\begin{IEEEbiography}[{\includegraphics[width=1in,height=1.25in,clip,keepaspectratio]{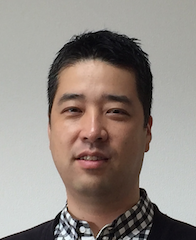}}]%
    {Naohisa Sakamoto} is an associate professor of the Graduate School of System Informatics at Kobe University. He received the Ph.D. degree in Graduate School of Engineering from Kyoto University in 2007. From 2001 to 2003, he worked in KGT (Kubota Graphics Technology) in Japan. His research interests include scientific visualization and visual analytics. He is a member of the IEEE Computer Society, the Visualization Society of Japan, and the Japan Society for Simulation Technology.
\end{IEEEbiography}
\vspace{-0.35in}

\begin{IEEEbiography}[{\includegraphics[width=1in,height=1.25in,clip,keepaspectratio]{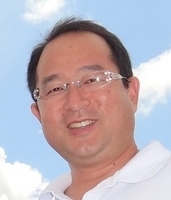}}]%
    {Jorji Nonaka} is a technical scientist at the Operations and Computer Technologies Division in RIKEN R-CCS, Japan. He received his M.Sc. degree in computer science from University of Brasilia, Brazil, and his Ph.D. degree in Informatics from Kyoto University, Japan. Before joining RIKEN, he worked at KGT Inc., Japan, and Federal University of Pernambuco (UFPE), Brazil. His current interests and activities include large data analysis and visualization, HPC usability, and energy efficiency improvement. 
\end{IEEEbiography}
\vspace{-0.35in}

\begin{IEEEbiography}[{\includegraphics[width=1in,height=1.25in,clip,keepaspectratio]{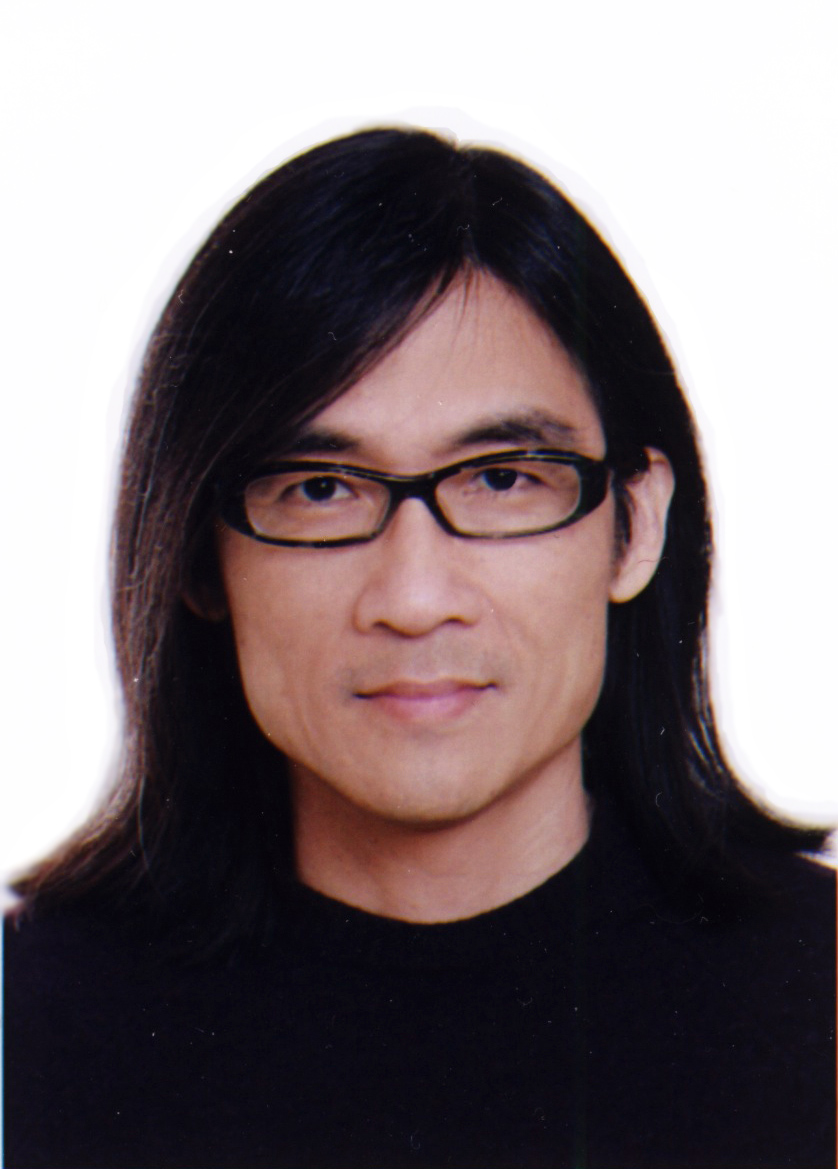}}]%
    {Kwan-Liu Ma} is a distinguished professor of computer science at the University of California, Davis. He received his PhD degree in CS from the University of Utah in 1993. His research is in the intersection of visualization, computer graphics, human-computer interaction, and high performance computing.  For his significant research accomplishments, Ma received special recognition, which includes being elected as IEEE Fellow (2012), recipient of the IEEE VGTC Visualization Technical Achievement Award (2013), and inducted to IEEE Visualization Academy (2019).
\end{IEEEbiography}



\end{document}